\documentclass[prl,aps,twocolumn,preprintnumbers,amsmath,amssymb,nofootinbib,superscriptaddress,notitlepage]{revtex4-1}
\usepackage{float}
\usepackage{epsfig}
\usepackage[utf8]{inputenc}
\usepackage{color}
\usepackage{epstopdf}
\usepackage[colorlinks=true,
linkcolor=blue,
breaklinks=true,
urlcolor=blue,
citecolor=blue]{hyperref}
\usepackage{diagbox}
\newcommand{\nn}{\nonumber}
\newcommand{\be}{\begin{equation}}
\newcommand{\ee}{\end{equation}}
\newcommand{\bea}{\begin{eqnarray}}
\newcommand{\eea}{\end{eqnarray}}
\newcommand{\ba}{\begin{array}}
\newcommand{\ea}{\end{array}}
\newcommand{\bi}{\begin{itemize}}
\newcommand{\ei}{\end{itemize}}

\newcommand{\mcl}{{\mathcal L}}

\newcommand{\difd}{\mathrm d}

\renewcommand{\vec}[1]{\mbox{\boldmath $#1 \!\!$ \unboldmath}}

\newcommand{\lf}{\left}
\newcommand{\rg}{\right}

\makeatletter
\newcommand{\EndMatterTitle}{%
  \par\addvspace{1.5ex}%
  \onecolumngrid@push
  \begingroup
    \centering\large\bfseries End Matter\par
  \endgroup
  \nobreak\@nobreaktrue
  \addvspace{1.5ex}%
  \onecolumngrid@pop
}
\makeatother

\newcommand{\beginendmatter}{%
  \EndMatterTitle
  \setcounter{equation}{0}%
  \renewcommand{\theequation}{A\arabic{equation}}%
}

\newcommand{\ucas}{\affiliation{University of Chinese Academy of Sciences, Beijing 100049, China}}

\newcommand{\keylab}{\affiliation{State Key Laboratory of Heavy Ion Science and Technology, Institute of Modern Physics, Chinese Academy of Sciences, Lanzhou 730000, China}}

\newcommand{\hebei}{\affiliation{Department of Physics, Hebei University, Baoding 071002, China}}

\newcommand{\nanjing}{\affiliation{Nanjing Normal University, Nanjing, Jiangsu 210023, China}}

\begin{document}

\title{Constraining DVCS Compton Form Factors \\ Using Lattice QCD informed Neural Network}

\author{Yuan-Yuan Huang}
\keylab
\nanjing
\hebei

\author{Xu Cao}\email{Corresponding author: caoxu@impcas.ac.cn}
\keylab
\ucas

%\author{Taifu Feng}\email{fengtf@hbu.edu.cn}
%\hebei

%\author{Kre\v{s}imir Kumeri\v{c}ki}\email{kkumer@phy.hr}
%\Zagreb

%\author{Yu Lu}\email{ylu@ucas.ac.cn}
%\ucas

\date{\today}

\begin{abstract}
  \rule{0ex}{3ex}
    The lattice QCD calculation of generalized form factors are exploited to determine the subtraction constants through all order dispersion relations of Deeply Virtual Compton Scattering (DVCS). The leading order relation is found to constrain significantly the real part of the Compton Form Factors (CFFs), and the higher order one reduces considerably both the real and imaginary part of CFFs in a global analysis of proton data.
    This is realized by a synthesis of the DVCS data and LQCD calculations within a neural network framework, whose architecture is specifically designed for a reliable extrapolation to unmeasured kinematic regime.
    By leveraging dispersion relations beyond leading order, our framework allows for adding higher moments of generalized parton distributions from LQCD into the extraction of CFFs from DVCS data.
\end{abstract}

\maketitle

%\section{Introduction}
{\it{Introduction.}}---
Achieving a quantitative description of the three dimensional structure of
the proton in terms of generalized parton distributions (GPDs) \cite{Muller:1994ses,Ji:1996ek} is one of the primary scientific goals of contemporary JLab and CERN facilities and the future electron-ion
colliders \cite{Accardi:2012qut,AbdulKhalek:2021gbh,Anderle:2021wcy}.
Among the exclusive processes relevant to GPDs, the deeply virtual Compton scattering (DVCS) is the most extensively studied both experimentally and theoretically \cite{Mueller:1998fv,Ji:1996nm,Radyushkin:1997ki}.
However, extraction GPDs directly from data is a defining challenge for the field of modern hadron physics
as they enter experimental observables through complex Compton form factors (CFFs) ~\cite{Diehl:2003ny,Belitsky:2005qn,Guidal:2013rya,Diehl:2023nmm,Boer:2025ixc,Alexandrou:2026jpd}.
The shadow GPDs highlights the issue of a complicated convolution of GPDs with partonic hard-scattering amplitudes \cite{Bertone:2021yyz,Moffat:2023svr}.
The determination of CFFs, more or less directly, from experimental observables constitutes an intermediate step towards the extraction of GPDs from the corresponding form factors.
The artificial neural network (NN) technique is demonstrated to be a flexible parameterization of CFFs, allowing for a considerable reduction of model bias \cite{Kumericki:2011rz,Moutarde:2019tqa,Cuic:2020iwt,CaleroDiaz:2025luc,Grigsby:2020auv,Almaeen:2024guo,Almaeen:2022imx,Adams:2024pxw,Hossen:2024qwo}.
Embodying the dispersion relation \cite{Polyakov:2002yz,Teryaev:2005uj,Anikin:2007yh,Polyakov:2018zvc} which connects real and imaginary part of CFFs into NN have already imposed additional constraints on CFFs \cite{Cuic:2020iwt}.

Complementary to experiment, lattice QCD offers a first-principles approach to the moments of nucleon GPDs with recent calculations now performed directly at the physical pion mass \cite{Bali:2018zgl,Alexandrou:2019ali,Hackett:2023rif}, in contrast to earlier work at unphysical pion masses \cite{LHPC:2007blg,Constantinou:2020hdm,HadStruc:2024rix}.
%x-moments of GPDs: form factors and generalized form factors (GFFs)
Recent advances in large-momentum effective theory \cite{Ji:2013dva,Ji:2020ect} and  have enabled access to skewness-dependent GPDs through lattice QCD calculations, whose moments have been compared with earlier lattice results obtained via the operator product expansion \cite{Lin:2020rxa,Alexandrou:2020zbe,Bhattacharya:2023ays,Dutrieux:2026grg}.
Despite their complementarity, LQCD results have not yet been combined with DVCS data to constrain CFFs, for two main reasons.
First, global DVCS analyses with reasonable uncertainty quantification remain at the CFF level, bypassing an explicit GPD parametrization; this approach requires knowledge of the full three-dimensional kinematic dependence of GPDs—namely on the parton momentum fraction $x$, skewness $\xi$, and momentum transfer $t$—which is hindered by the deconvolution problem.
A recent attempt along these direction is the application of neural networks to the representation of GPDs \cite{Watkins:2025apc,Xu:2026lko,Riberdy:2023awf,Dutrieux:2021wll}.
Second, a flexible, minimally model-dependent parametrization of GPDs that is valid across all skewness values is required to enable a direct comparison between empirical extractions from exclusive processes and LQCD calculations \cite{Mamo:2024jwp,Dotson:2025omi,Panjsheeri:2025vpa}.
%from $t$-Channel String Exchange in AdS Spaces
Several GPD models have already been used in global fits to DVCS data~\cite{Kumericki:2007sa,Kumericki:2009uq,Moutarde:2018kwr,Cuic:2023mki,Guo:2023ahv}, some of which integrate recent lattice QCD results \cite{Guo:2025muf}.

Recently the flavor decomposition of the gravitational form factors (GFFs) of proton is obtained from LQCD close-to-physical value of the pion mass \cite{Hackett:2023rif}.
They are consistent with one of its quark GFFs from experimental measurements of DVCS \cite{Burkert:2018bqq,Burkert:2023wzr,Kumericki:2019ddg} and gluonic GFFs from threshold charmonium $J/\psi$  photoproduction  \cite{Duran:2022xag,Guo:2025jiz}.
The mechanical properties of the proton are encoded in its gravitational form factors (GFFs), among which the Polyakov–Weiss $D$-term plays a central role in characterizing the internal forces and pressure distributions \cite{Polyakov:1999gs,Ji:2025qax}.
The dispersion relations allow for a combination of GFFs from LQCD and DVCS data if overlooking the shadow $D$-term \cite{Dutrieux:2021nlz,Dutrieux:2024bgc} under some reasonable approximation at the moment.
At first glance GFFs would indeed provide valuable input of global analysis of CFFs by adding more theoretical consideration into the global fitting of DVCS data.
However, this extraction procedure is also complicated by the integration in the dispersion relation, therefore involving a careful consideration of systematic uncertainties from extrapolation.
In practice, a synthesis of machine learning techniques and physics-driven designs is reforming the way we address a full three-dimensional picture of the nucleon structure,
in which accurate physical properties are extracted from complex data sets.
%neural networks, due to their exceptional flexibility and fitting capabilities, have become powerful tools for characterizing the three-dimensional distribution of CFFs.
By a careful design of architecture and scrutinization of hyperparameters,
a LQCD-informed neural network leveraging  dispersion relations of DVCS and possessing excellent extrapolation performance is suitable for the purpose of a global analysis of DVCS data and LQCD calculation of GPDs moments.
From another perspective, this will offer a systematic check of the consistency between DVCS data and LQCD moment approaches to the GPDs in a model independent manner.

%\section{Basics of dispersion relations}
{\it{Basics of dispersion relations.}}---
The present numerical attempt remains at the determination of leading-twist
Compton form factors (CFFs $\mathcal{H}$, $\mathcal{E}$, $\tilde{\mathcal{H}}$ and $\tilde{\mathcal{E}}$) without quark flavours ($q$) separated at the
level of the leading order (LO) of the perturbation theory \cite{Cuic:2023mki}.
The data of DVCS at JLab, COMPASS, HERMES, and HERA (H1 and ZEUS) are truncated with the kinematic cuts $Q^2 > 1.5~\text{GeV}^2,  -{t}/{Q^2} < 0.2$
with the purpose to restrict the phase space covered by experiment to the deep virtual region.
The whole framework is apparently ready to extend at least to the flavour separation
at LO level, and potentially advancing to the gluon contribution appearing at the NLO level in the future.

A general expression for the $n$-times subtracted dispersion relation at any order of perturbation theory has been established as \cite{Diehl:2007jb,Dutrieux:2024bgc,Martinez-Fernandez:2025jvk,Martinez-Fernandez:2025rcg}
\begin{multline} \label{eq:DR}
\mathfrak{Re} \mathcal{H}^q(\xi, t) =  \sum_{j \, \text{even}}^n h^q_j \xi^{-j}
+ \\ \frac{1}{\pi} \, \text{P.V.} \int_0^1 \difd x \left[ \frac{1}{\xi - x} - \frac{(-1)^n}{\xi + x} \right] \lf(\frac{x}{\xi}\rg)^n \mathfrak{Im} \mathcal{H}^q(x, t)
\end{multline}
by omitting the evolution scale $\mu^2 = Q^2$.
Above formula is exactly identical to the usual LO formula when $n$ = 0 (or equally $n$ = 1) \cite{Teryaev:2005uj}.
The kinematical variables are defined as squared momentum transfers from the lepton $Q^2$ and to the nucleon $t$, the skewness $\xi = x_B/(2-x_B)$ with Bjorken $x_B$.
The $x + \xi$ and $x - \xi$ are the longitudinal
momentum fractions of the quark before and after
radiating the real photon, respectively.
The subtraction constants
can be written in terms of an infinite linear combination of the generalised form factors $\bar{D}^q_{m}(t)$, $A^q_{m,2j}(t)$ and $B^q_{m,2j}(t)$, which is related to higher Mellin moments of GPDs due to their polynomiality properties \cite{Dutrieux:2024bgc}.
If taking into account kinematic
power corrections and truncating the series to first moment of GPDs \cite{Martinez-Fernandez:2025jvk,Martinez-Fernandez:2025rcg,Braun:2014sta,Braun:2022qly,Braun:2025xlp},
%the two first elements neglecting the terms of $n \geq 2$.
%the LO subtracted constants $h_{0}$ is:
%\begin{widetext}
\bea \label{eq:subLO}
h_0^q(t) &\simeq& 5 \bar{D}_1^q(t) \lf(1 - \frac{t}{3 Q^2} \rg) \nn \\ &-& \frac{4 M^2}{Q^2} c_0^{11} \lf[ (1- \frac{t}{4 M^2}) A^q(t) + \frac{t}{2 M^2} J^q(t) \rg] %\simeq 5 \bar{D}_1^q(t)
\eea
%\end{widetext}
with $c_0^{11} =0.8645$.
%At the LO approximation $h_0^q(t) \simeq 4 d_{1}^q = 5 \bar{D}_1^q(t)$ means that $\bar{D}_1^q(t)$ from LQCD calculation could be used to .
%\begin{multline} \label{eq:DRLO}
%\mathfrak{Re} \mathcal{H}^q(\xi, t) = h_0^q(t) \\
%+ \frac{1}{\pi} \, \text{P.V.} \int_0^1 \difd x \left( \frac{1}{\xi - x} - \frac{1}{\xi + x} \right) \mathfrak{Im} \mathcal{H}^q(x, t)
%\end{multline}
Here $h_0^q(t)$ is a LO subtraction constant in $\xi$, which is up to an opposite sign the same for $\mathcal{H}^q$ and $\mathcal{E}^q$ (e.g. $e_0^q(t) = - h_0^q(t)$), and is zero for $\tilde{\mathcal{H}}^q$ and $\tilde{\mathcal{E}}^q$.
$\bar{D}_1^q(t)$, $A^q(t) = A^q_{1,0}(t)$, and $J^q(t) = (A^q_{1,0}(t) + B^q_{1,0})/2$ are GFFs defined through the nucleon matrix elements of the traceless, symmetric energy-momentum tensor (EMT) \cite{Ji:1996ek} and related to first moments of GPDs.
In above relation the dominant contribution is from the $\bar{D}_1^q(t)$ in terms of operator $(\Delta^\mu\Delta^\nu- g^{\mu\nu} \Delta^2)/4M_N^2$ with momentum transfer between the incoming ($p$) and outgoing $p'$ nucleons $\Delta^\mu = (p' -p)^\mu$ and its square $t = \Delta^2$.
This is resulted from the fact that $h_0^q(t)$ and $\bar{D}_1^q(t)$ both are given by integrals over the $D$-term (Drucker term) form factors of GPDs, differing only in their integration kernels:
\bea \label{eq:GFFD1q}
\bar{D}_1^q(t) &=& \int_{-1}^{1} z D^q(z,t) \difd z = \frac{4}{5}\, d_1^q \,, \\
h_0^q(t) &=& 2 \int_{-1}^{1} \frac{D^q(z,t)}{1-z} \difd z = 4 \sum_{k=0}^{+\infty} d_{2k+1}^q \,, \label{eq:h0q}
\eea
%$\Delta^q(t) &\equiv& h_0^q(t)$
with a parametrization of the $D^q(z,t)$ through an expansion in Gegenbauer moments $d_{i}^q$.
The deconvolution problem of $D$-term is neglected by dropping higher order tems in Eq. \eqref{eq:h0q}.

The higher order (HO) subtraction with $i \geq 1$ reads \cite{Dutrieux:2024bgc}
\be \label{eq:subHO}
h_{2i}^q(t) = \frac{2}{\pi} \int_0^1  \mathfrak{Im} \mathcal{H}^q(\xi',t) (\xi')^{2i-1} \difd \xi'
\ee
with similar definition of $e_{2i}^q$ relating to CFF $\mathcal{E}^q$.
They are connected to the generalized form factors through
%\begin{widetext}
\bea
h_2^q(t) &\simeq& c_0^{01} A^q(t) - \frac{4 t }{Q^2} c_0^{11} \lf[ A^q(t) - J^q(t) \rg] \,,  %- \frac{8 M^2 -t }{2 Q^2} c_0^{13} A_{3,0} - \frac{t}{2 Q^2} c_0^{13} B_{3,0} \simeq c_0^{01} A^q(t)
 \\ h_2^q(t) &+& e_2^q(t) \simeq 4 \lf(c_0^{01} - \frac{2 t}{Q^2} c_0^{11} \rg) J^q(t) \,,  %- \frac{4 M^2 -t }{Q^2} c_0^{13} (A_{3,0} + B_{3,0}) \simeq 4 c_0^{01} J^q(t)
\\ h_4^q(t) &\simeq& \lf(\frac{1}{6} c_0^{03} - \frac{3 t }{2 Q^2} c_0^{13} \rg)  A_{3,0} +  \frac{2 t }{3 Q^2} c_0^{13} B_{3,0} %\qquad %- \frac{6 M^2 -t }{12 Q^2} c_0^{15} A_{5,0} - \frac{t }{24 Q^2} c_0^{15} B_{5,0} \simeq \frac{1}{6} c_0^{03} A_{3,0}
\\ h_4^q(t) &+& e_4^q(t) \simeq \frac{1 }{3} \lf( c_0^{03} - \frac{t}{ Q^2} c_0^{13} \rg)(  A_{3,0} + B_{3,0}) \,,  %- \frac{4 M^2 -t }{6 Q^2} c_0^{15} (A_{5,0} + B_{5,0}) \simeq \frac{1 }{3} c_0^{03} (  A_{3,0} + B_{3,0})
\eea
%\end{widetext}
with $c_0^{01} = 2 - 1.251 t/Q^2$, $c_0^{03} = 12 - 6.277 t/Q^2$ and $c_0^{13} = 1.9595$.
At present the subtractions relating to $\tilde{\mathcal{H}}^q$ and $\tilde{\mathcal{E}}^q$ are not yet considered.

Above all order dispersion relation with arbitrary
subtraction offer a framework of exploiting the lattice computations of the GFFs and Mellin moments of GPDs \cite{LHPC:2007blg,Bhattacharya:2023ays,Hackett:2023rif} as input to constrain the CFFs of DVCS.
The analysis herein is confined to the CFFs (and subtraction as well) of a summation over quark flavours because neutron DVCS measurement is of limited kinematic range and precision.
Only LO and NLO subtraction dispersion relation are enabled in the global fit because the state-of-art LQCD calculation of high Mellin moments of GPDs is at a pion mass of 260 MeV without including disconnected contributions. Alternatively, higher order subtractions $h_4$ and $e_4$ with uncertainties are given as our prediction in comparison of those reconstructed from the GFFs calculated within LQCD with a pion mass of 260 MeV \cite{Bhattacharya:2023ays}.
%\red{The $A_{3,0}$ and $B_{3,0}$ are with a different notation $A_{4,0}$ and $B_{4,0}$ therein}.
%\cite{LHPC:2007blg} pion masses as low as 350 MeV. \cite{Bhattacharya:2023ays} 260 MeV. \cite{Hackett:2023rif}

\begin{figure}
    \centering
    \includegraphics[width=\linewidth]{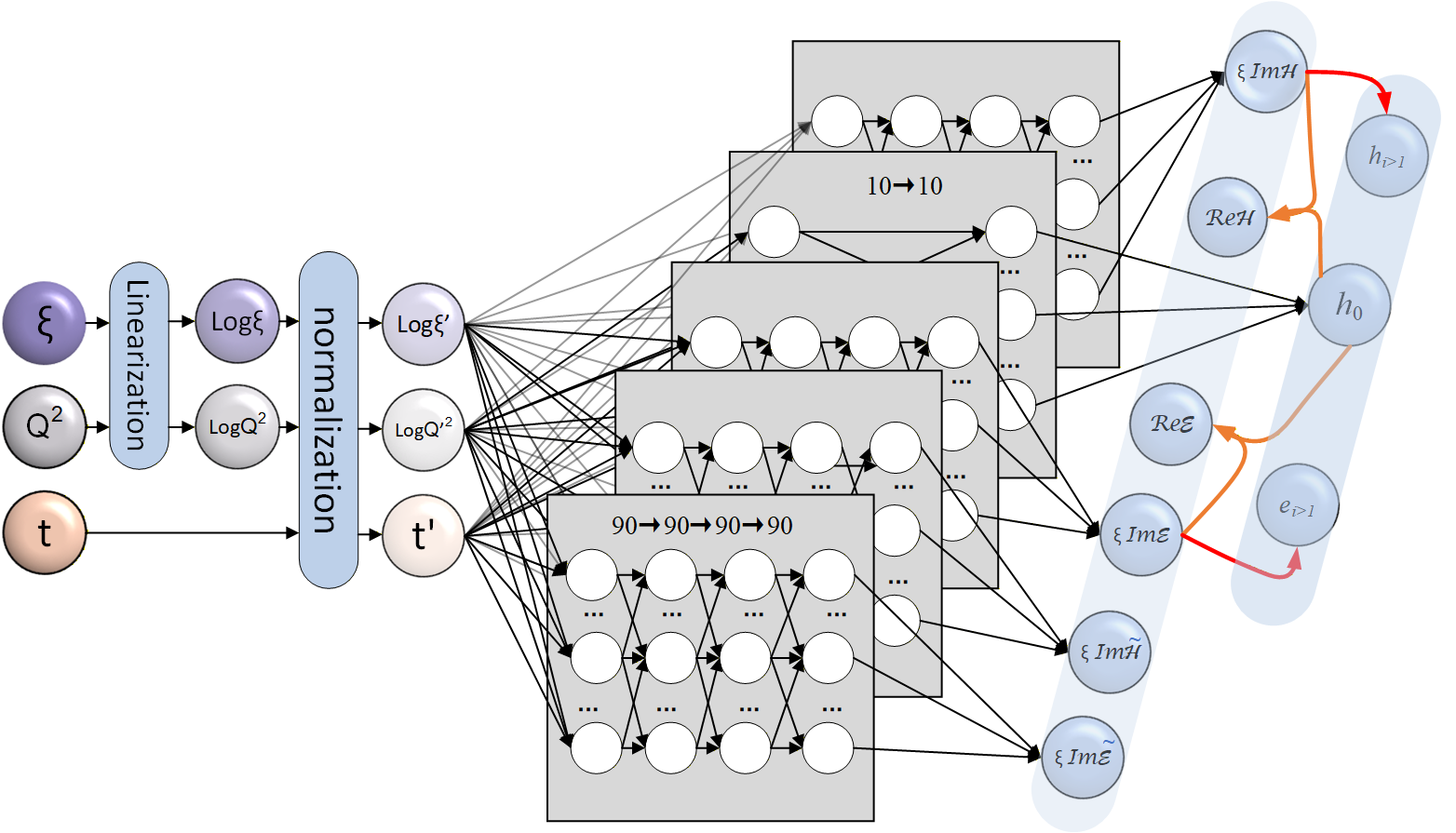}
    \caption{The neural network parameterization of the imaginary part of CFFs and LO subtraction constant. A single network is made out of three neurons in the input layer ($\xi$, $t$, $Q^2$) after linearization and normalization to give one output neuron, with each of four hidden layers processes ninety neurons (3 $\to 90 \to 90 \to 90 \to 90 \to$ 1) for CFFs and ten neurons (2 $\to 10 \to 10 \to$ 1) for $h_0$. The orange and red arrows label the LO and HO dispersion relations, respectively.}
    \label{fig:artNN}
\end{figure}

%\section{LQCD-informed neural network}
{\it{LQCD-informed neural network.}}---
The neural network parametrization of CFFs is proposed to relieve the model rigidity through different design of architectures \cite{Moutarde:2019tqa,Grigsby:2020auv,Almaeen:2024guo,Adams:2024pxw,Le:2025swl}, since its first realization in the field \cite{Kumericki:2011rz}.
Fig. \ref{fig:artNN} shows the architecture of neural network which represents the imaginary part of the CFFs $\mathfrak{Im} \mathcal{H}^q$ and $\mathfrak{Im} \mathcal{E}^q$ (and similarly $\mathfrak{Im} \tilde{\mathcal{H}}^q$ and $\mathfrak{Im} \tilde{\mathcal{E}}^q$) taking $\xi$, $t$ and ${Q^2}$ as input layer, and the LO subtraction constant $h_0(t)$ taking only $t$ and ${Q^2}$ as input layer.
The linearization of the kinematic variables $\xi$ and $Q^2$ is realized with a logarithmic projection.
Subsequently the experimental data of DVCS are normalized to conform to a standard Gaussian distribution in terms of the the kinematic space $\vec{x} = (\xi, t, Q^2)$ before entering the networks.
This preprocess ensures the stability and efficiency of neural network training
by mitigating vanishing or exploding gradients, accelerating convergence during training, and improving the overall generalization capability of the model. Additionally, it guarantees that the network treats all features equally, preventing any single feature from dominating the learning process due to its scale or range.
%with as three input neurons are of considering their covering a broad span of logarithmic scale.
%and sequential normalization $\xi$, $t$ and $Q^2$, to speed the optimization procedure.
%a careful regularization to avoid over-fitting and their training is more time-consuming
Real parts of a given CFF are then properly obtained using LO dispersive relation in Eq. \eqref{eq:DR}, and higher order subtractions are reconstructed through Eq. \eqref{eq:subHO}.
Afterwards DVCS observables are calculated within \texttt{Gepard} framework \cite{Cuic:2020iwt}.

%random division of the dataset into training and testing subsets. To prevent data leakage and overly optimistic model evaluation, we randomly split the dataset into a $70\%$ training set and a $30\%$ test set.
%The pretreatment normalization parameters are computed solely from the training data and then applied consistently to both sets.

The neural networks are trained with the back-propagation algorithm using Adam (Adaptive moment estimation) optimizer by minimizing the uncertainty-weighted Huber loss function \cite{Grigsby:2020auv,Almaeen:2024guo,Almaeen:2022imx,Adams:2024pxw,Hossen:2024qwo}:
\be \label{eq:loss}
\mcl_\theta^{\text{Huber}}(\vec{x},\sigma) = \mcl_\theta^{\text{Huber}}(\vec{x},\sigma_{\text{DVCS}}) + \lambda \mcl_\theta^{\text{Huber}}(\vec{x},\sigma_{\text{LQCD}})
\ee
where $\theta$ is the set of all the entries in the
transformation weights and biases of the hidden neurons, and
$\lambda$ is a weight that balances two terms of the DVCS data and LQCD calculation.
%The Huber loss is Mean Squared Error (MSE) function that is quadratic in the absolute error of our networks' predictions.
%when that error is less than a hyperparemter $\delta$, and linear otherwise \cite{Grigsby:2020auv}.
%It is the second term which imposes the LQCD constraint on the learnt function, and focuses the optimization over
The whole data collection is randomly divided into two independent sub-samples, training sample and test sample (70 : 30), for early stopping.
The activation function for neurons in the hidden layers is
\bea \label{eq:activation}
    f(x) &=&
   \begin{cases}
     0 \, & \mbox{for } \, x < 0 \\
     e^x-1 \, & \mbox{for } \, x > 0
   \end{cases}
\eea
which makes sure a reasonable extrapolating out of the regime constrained by measurements, leading to an appropriate assessment of the systematic uncertainties of integration over $x$ in its whole range in Eq. \eqref{eq:DR}.

Our implementation leverages \textit{PyTorch} for efficient processing of multi-dimensional data, enforcing a fast global analysis of DVCS data and LQCD constraints of subtraction constants.
A large number of replicas with bootstrap method are rapidly generated to propagate the experimental uncertainties to extracted quantities, a sample size 100 found to be sufficient for achieving stable results.
The hyperparameters of neural network, the number of hidden layers and neurons per layer, are validated by the pseudo-data of future electron-ion colliders \cite{Cao:2023wyz,Huang:2025wdq}.
%, see Appendex \ref{apdx:closure}.

\begin{figure}[htb]
    \centering
    \includegraphics[width=\linewidth]{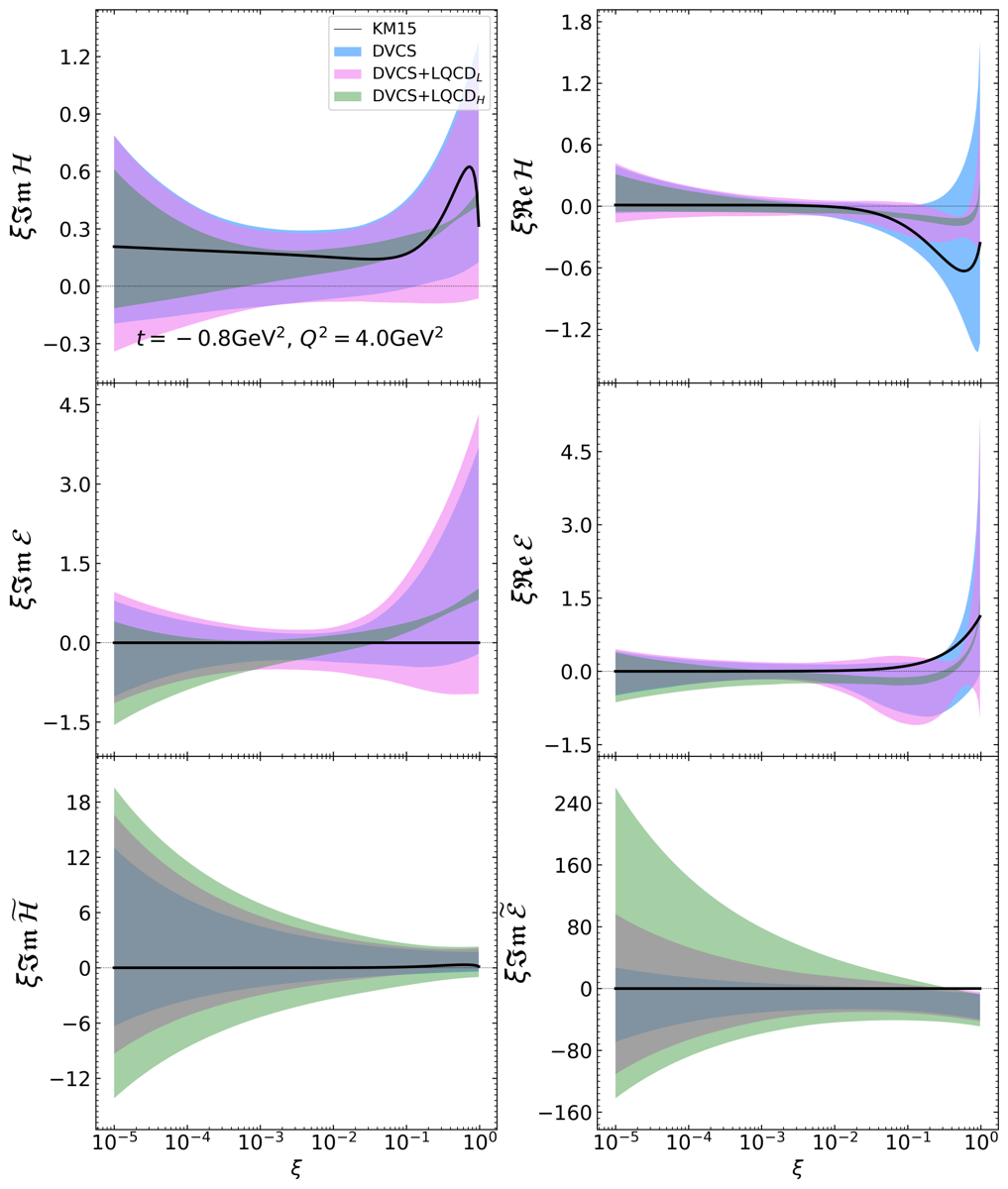}
    \caption{Imaginary and real parts of the CFF as a function of $\xi$ for $t = -0.8$ GeV$^2$ and $Q^2 = 4$ GeV$^2$ in comparison of KM15 model (black line).
    Blue band: those without incorporating subtractions; magenta bands: those including only LO subtractions; green band: those including HO subtractions. }
    \label{fig:CFFxi}
\end{figure}
%xcff_log_xi_t08_Q24_KM15.pdf
% before (blue bands) and after considering the (LO: magenta bands, HO: green band)  LQCD constraints.

\begin{figure}[htb]
    \centering
    \includegraphics[width=\linewidth]{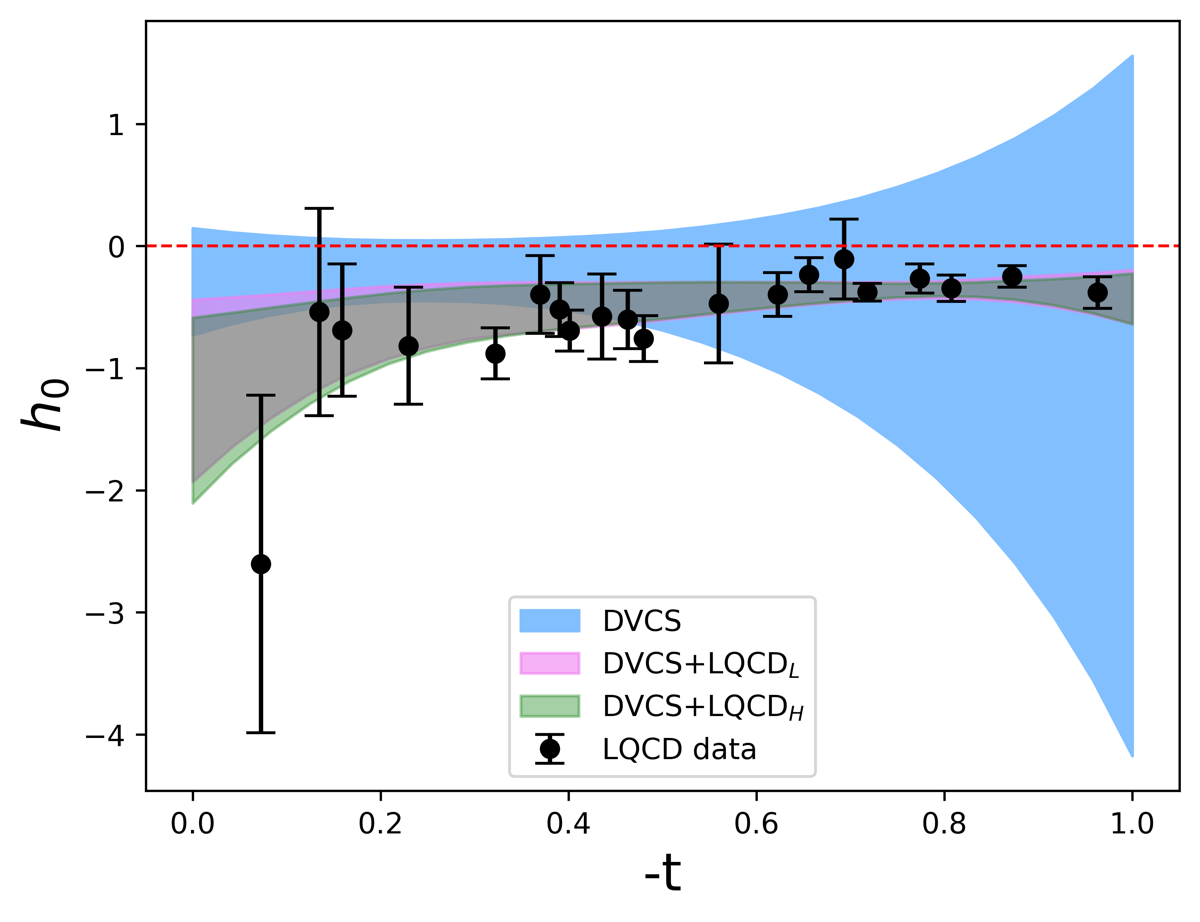}
    \caption{The LO subtraction constant of proton $h_0(t)$ as a function of $-t$ for $Q^2=4.0$ GeV$^2$ in comparison of LQCD calculation in the $\bar{MS}$ scheme at scale $\mu =$ 2 GeV. The solid points are reconstructed from the flavor separated $\bar{D}_1^q(t)$ with correlation matrix within LQCD calculations \cite{Hackett:2023rif}. }
    \label{fig:subh0_4}
\end{figure}
%h0_t_Q24.pdf
% The central values at $-t=0$ is labeled by the black numeric.

\begin{figure}[htb]
    \centering
    \includegraphics[width=0.48\linewidth]{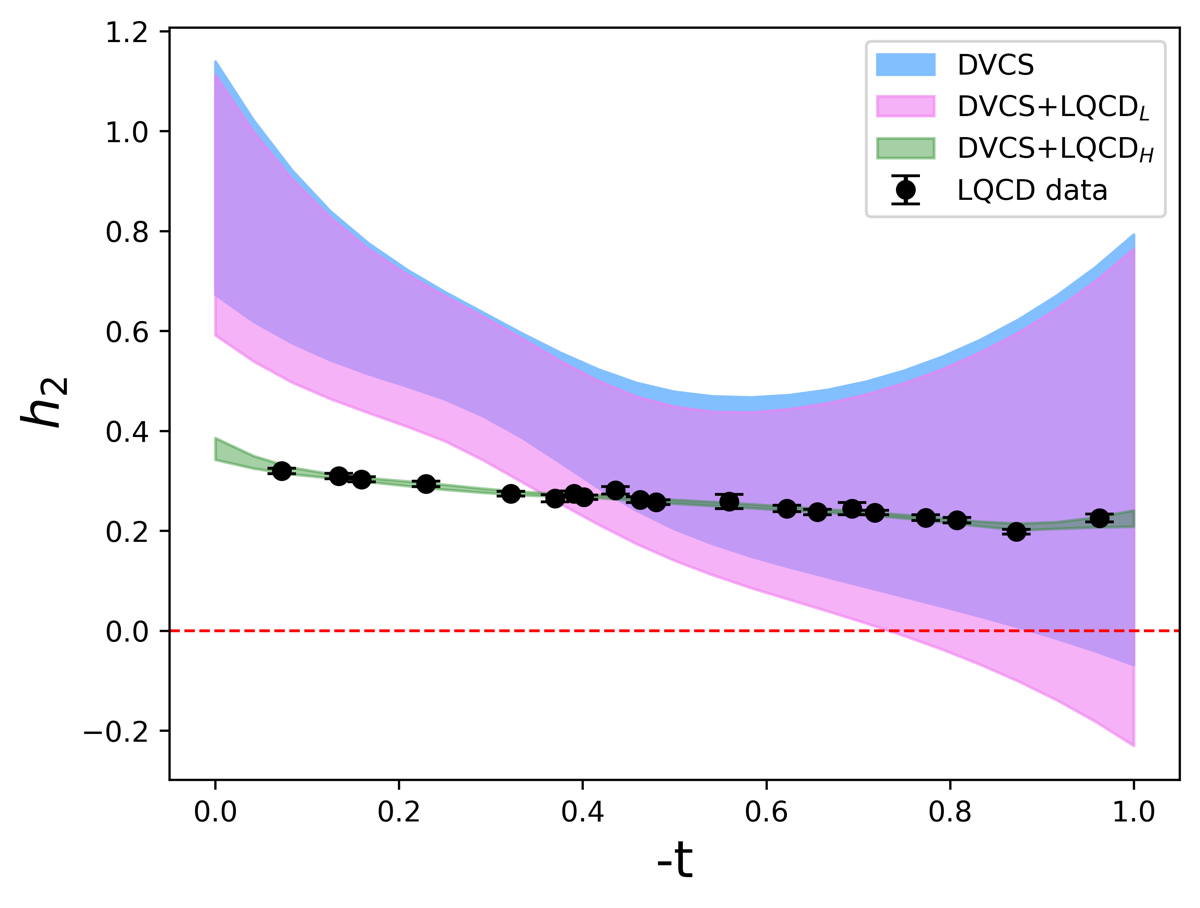}
    \includegraphics[width=0.48\linewidth]{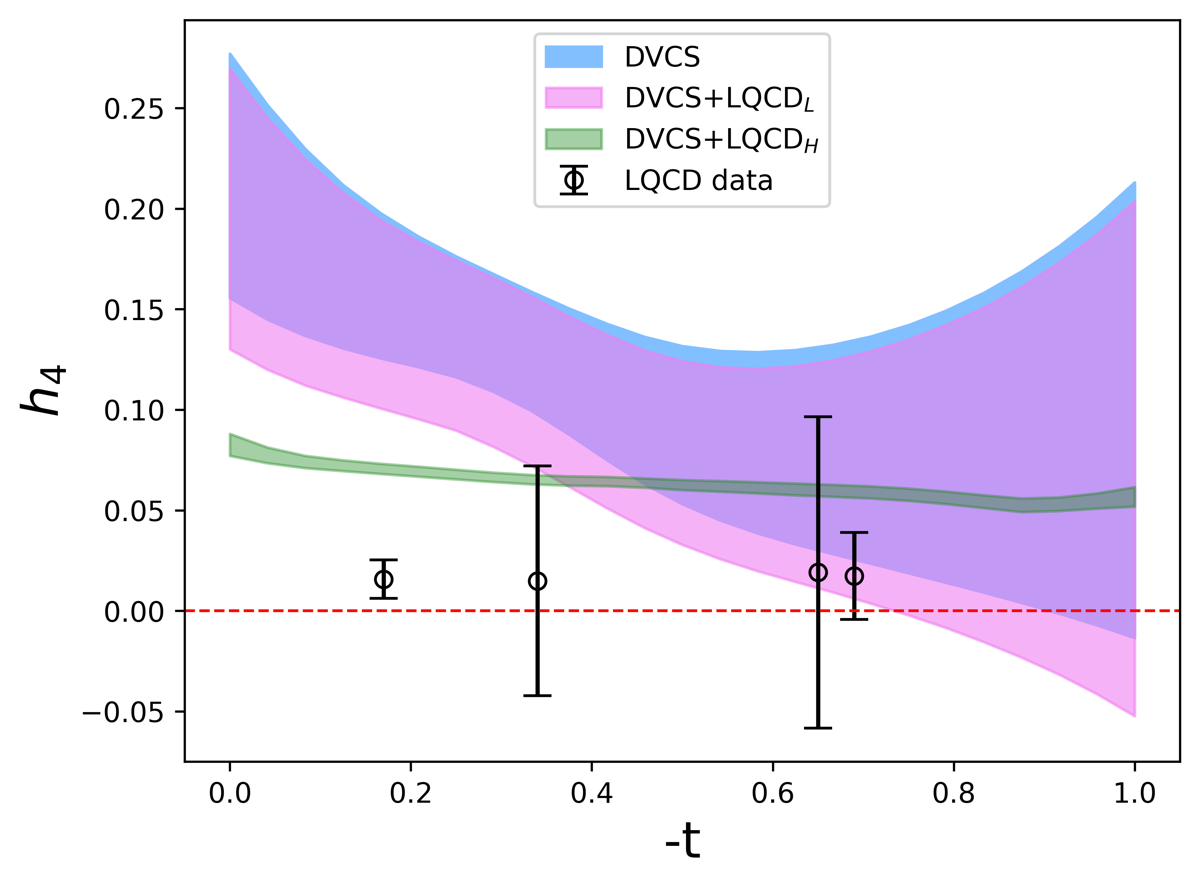}
    \includegraphics[width=0.48\linewidth]{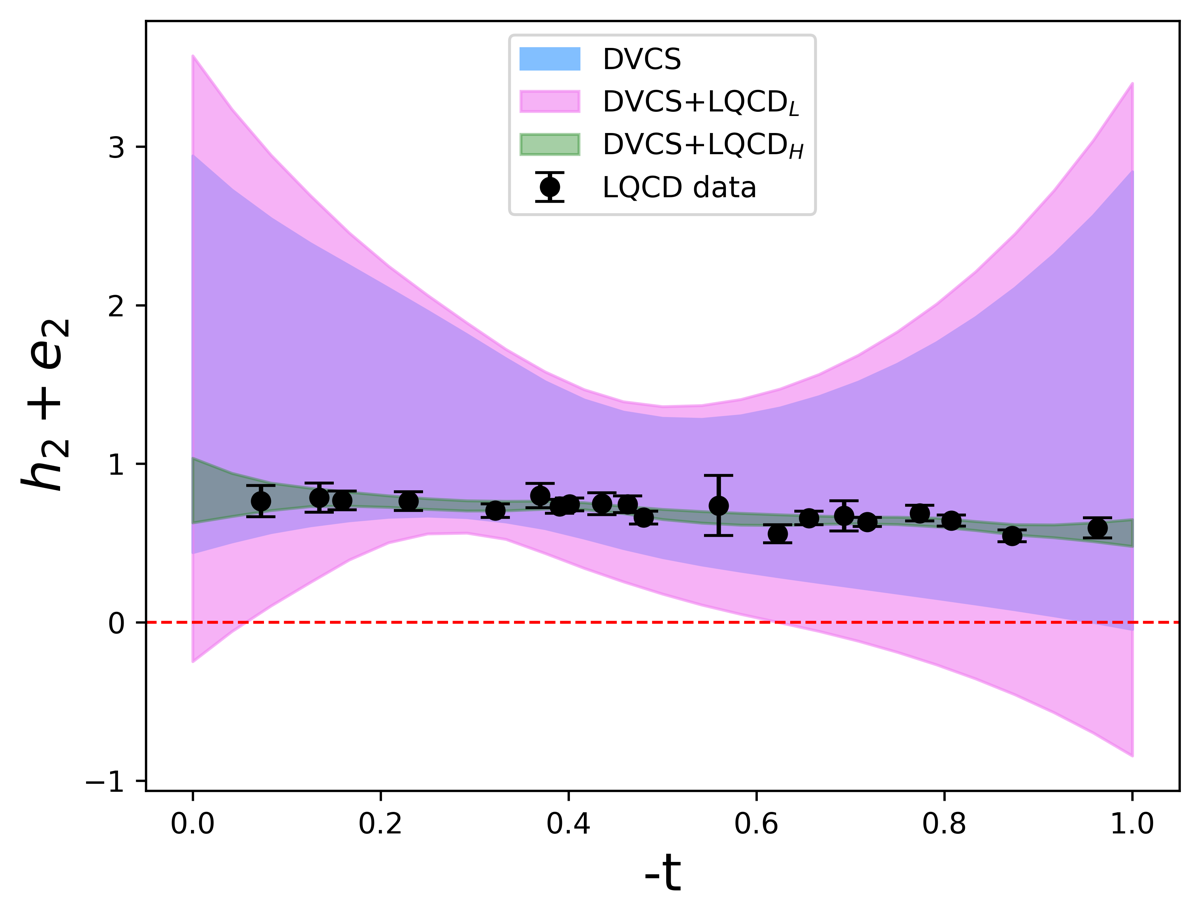}
    \includegraphics[width=0.48\linewidth]{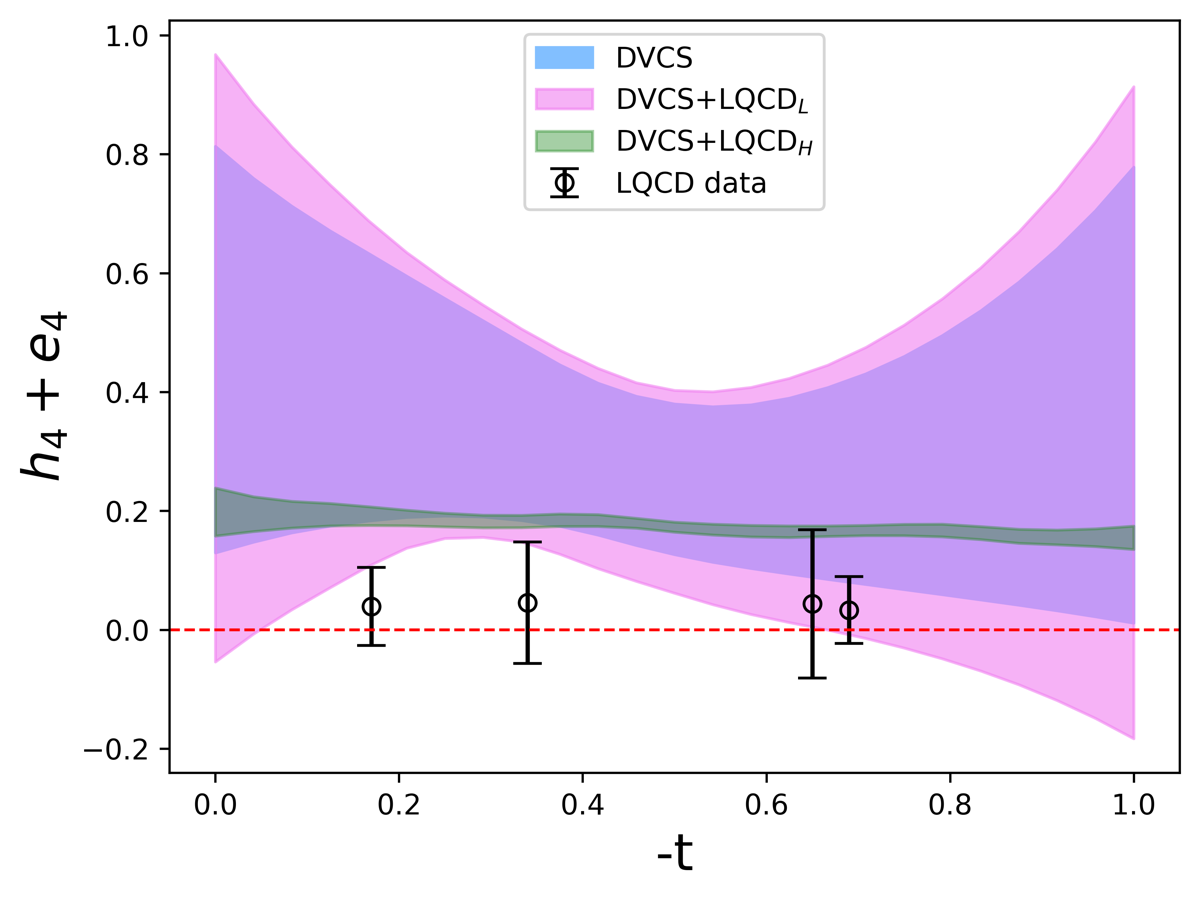}
    \caption{Higher order subtractions $h_2$, $h_2+e_2$, $h_4$, and $h_4+e_4$ with uncertainties as a function of $-t$ for $Q^2=4.0$ GeV$^2$.The solid points of $h_2$ and $h_2+e_2$ are constructed from the flavor separated generalized form factors with correlation matrix within LQCD \cite{Hackett:2023rif}. The empty points of $h_4$ and $h_4+e_4$ are constructed with a naive error propagation from generalized form factors in LQCD \cite{Bhattacharya:2023ays} and not included in the global fit.}
    \label{fig:subh2_4}
\end{figure}
%h2_t_Q24.pdf
%h4_t_Q24.pdf
%h2pe2_t_Q24.pdf
%h4pe4_t_Q24.pdf

% Performance

%\section{Impact of LQCD constraints}
{\it{Impact of LQCD constraints.}}---
It is confirmed that present data of DVCS are not sensitive at all to $\mathfrak{Re}\tilde{\mathcal{H}}$ and $\mathfrak{Re}\tilde{\mathcal{E}}$  \cite{Cuic:2020iwt,Moutarde:2019tqa}.
The neural network fits well the available data of proton DVCS by using just the remaining six CFFs.
The fit qualities assessed by $\chi^2$ per single data point remain reasonably stable before and after including the constraints of subtraction from LQCD calculations, indicating the consistency between them.

% Compton form factors

%\red{PARTONS:} The data provide the best constraints on $\mathfrak{Im}{\mathcal{H}}$, and some on $\mathfrak{Re}{\mathcal{H}}$, $\mathfrak{Im}\tilde{\mathcal{H}}$ and $\mathfrak{Re}\tilde{\mathcal{E}}$. Other CFFs are poorly constrained by the available data, in particular $\mathcal{E}$ related to GPD $E$.

An extraction of six CFFs without the inclusion of LQCD constraints in Fig. \ref{fig:CFFxi} shows that the DVCS data provide the best constraints on $\mathfrak{Im}{\mathcal{H}}$, and some on $\mathfrak{Re}{\mathcal{H}}$.
The uncertainties
are increasingly large in regime sparsely covered by data,
preventing successfully from introducing a bias on the integration in the dispersion relation.
The CFFs in our fit is in broad agreement with the results of the (also DR-constrained) fit of previous \texttt{Gepard}, but is supported by more solid validation \cite{Huang:2025wdq}.

The results shows significant constraints on both the $\mathfrak{Re}\mathcal{H}$ and $\mathfrak{Re}\mathcal{E}$ within the range $[10^{-1}, 10^0)$ of $\xi$ after incorporating LQCD constraints via LO dispersive relation.
Specifically, this additional constraint is manifested by a noticeable narrowing of the uncertainty bands, in particular for $\xi$ approaching to 1.0.
The behaviour of CFFs as as a function of $\xi$ prefers a positive $\mathfrak{Re}\mathcal{H}$ close to $\xi = 1.0$, but indicates no tendency of $\mathfrak{Re}\mathcal{E}$.
The two bands overlap to each other, demonstrating the consistency between two fits with and without the inclusion of LQCD constraints.
The $Q^2$ and $t$ dependence of CFFs $\mathfrak{Re}{\mathcal{H}}$ and $\mathfrak{Re}{\mathcal{E}}$ are also sensitive to the LQCD constraints, in particular for the domain of large $Q^2$ and $t$, see the End Matter.
%Appendix \ref{apdx:closure}.

The LO subtraction constant of proton $h_0(t)$ as a function of $-t$ for $Q^2=4.0$ GeV$^2$ in Fig. \ref{fig:subh0_4} shows that DVCS data alone do not give much constrain to the $h_0(t)$, in particular for $h_0(t=0) = - 0.40 \pm 0.50$.
Our findings are consistent with the claim in ref. \cite{Kumericki:2019ddg} and
observation of PARTONS \cite{Moutarde:2019tqa}.
The LQCD calculation of $h_0(t) = \sum_q e_q^2 h_0^q(t)$ in Eq. \eqref{eq:subLO} could be constructed from the flavor separated $\bar{D}_1^q(t)$ with correlation matrix at a pion mass of 170 MeV in the $\bar{MS}$ scheme at scale $\mu =$ 2 GeV \cite{Hackett:2023rif}, as shown by the solid points in Fig. \ref{fig:subh0_4}.
The pion mass dependence is weak in this range of pion mass as shown in the dispersive evaluation \cite{Pasquini:2014vua,Cao:2024zlf,Cao:2025dkv}.

Our neural network describes quite well those LQCD calculations, giving $h_0(t=0) = - 1.28 \pm 0.70$, slightly favoring a negative value.
If assuming equal contribution of up and down quark and negligible strange quark to $D$-term as preferred by the current precision of LQCD calculation \cite{Hackett:2023rif}, our value $D^{u+d} = -0.92 \pm 0.50$ (and $-0.29 \pm 0.36$ without LQCD constraints) determined from global proton DVCS data is confronting with positivity bound $D \leq -0.20 \pm 0.02$ \cite{Gegelia:2021wnj}, $-3.35 \pm 0.58$ in $z$-expansion of LQCD calculation \cite{Hackett:2023rif} and $-3.38^{+ 0.34}_{- 0.35}$ in dispersive determination \cite{Cao:2024zlf,Cao:2025dkv}, indicating significant role of gluon \cite{Hackett:2023rif}.
The afore assumption could be loosen if the data of neutron DVCS are precise enough and of broad kenimatic coverage for a meaningful flavor separation \cite{Cuic:2023mki}.

\begin{figure}[H]
    \centering
    \includegraphics[width=0.98\linewidth]{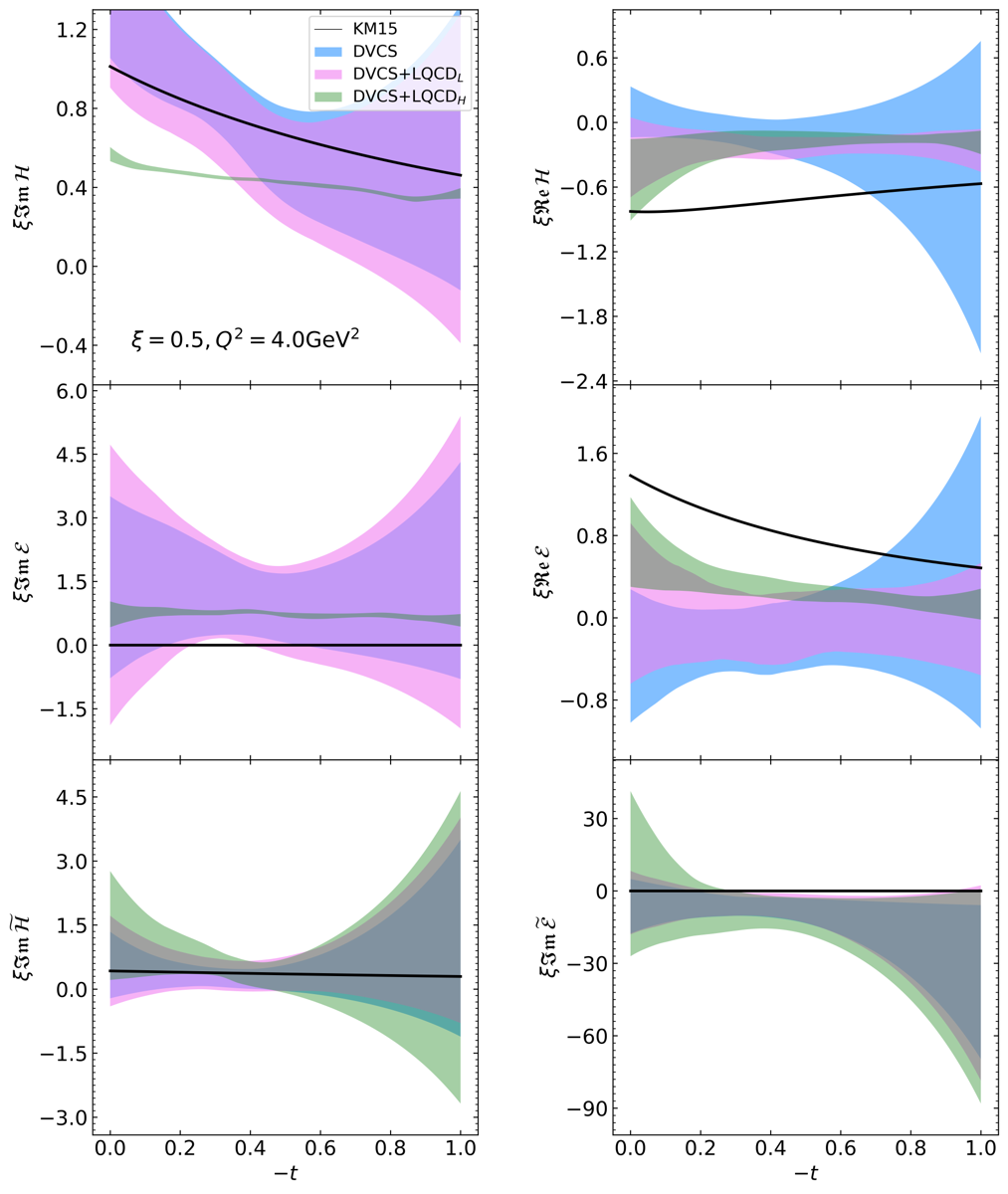}
    \caption{Imaginary and real parts of the CFFs as a function of $-t$ for $\xi=0.5$ and $Q^2=4$ GeV$^2$}
    \label{fig:CFFt}
\end{figure}

FIg. \ref{fig:subh2_4} gives the predictions of the higher order subtractions with uncertainties. $h_2$ and $h_4$ are clearly predicted to be positive and notably to decrease for $-t < 1.0$ GeV$^2$ by the neural network fits.
$h_2+e_2$ and $h_4+e_4$ are of little dependence on $-t$ and also positive with large uncertainties, showing that present LQCD contraints do not give more information on those subtractions.
%but consistent with zero
%in the considered $t$ range at present.  As expected the absolute central values of subtractions decrease when going to higher order.
The predictions are consistent with the subtractions reconstructed from GFFs in LQCD over a wide $-t$ range.
However, it should be noted that all considered subtractions as reconstructed from LQCD GFFs \cite{Hackett:2023rif} depend weakly on $-t$, and $h_2$ and $h_4$ show some deviation from the LO fits for $-t<0.4$ GeV$^2$.
After further including those information of $h_2$ and $h_2 +e_2$ from LQCD calculations not only the uncertainties of real part of CFFs but also the imaginary part reduce considerably as shown in Fig. \ref{fig:CFFxi}.
As a result of above mentioned deviation the central values of $\mathfrak{Im}{\mathcal{H}}$ for $-t<0.4$ GeV$^2$ shift downward relative to previous fits as shown in Fig \ref{fig:CFFt} in End Matter.% Appendix.
All the included subtractions are well described in the global fit, and the $h_0$ remains stable whether the LO or NLO dispersion relation is enabled as shown in Fig. \ref{fig:subh0_4}.
$h_0(t=0) = - 1.07 \pm 0.72$, or $D^{u+d} = -0.77 \pm 0.55$ is consistent with that in LO dispersion relation.
Small positive values are predicted for $h_4$ and $h_4+e_4$, confronting with those reconstructed from generalized form factors in LQCD calculations at large pion mass \cite{Bhattacharya:2023ays}.

Therefore our results validate the input of LQCD in providing additional information on both real and imaginary part of CFFs, especially in large $\xi$, $-t$ and $Q^2$ regions.
The CFFs $\tilde{\mathcal{H}}$ and $\tilde{\mathcal{E}}$ seems to be not affected by the LQCD constraints. However this shall be instead considered as a verification of the consistency of our neural networks fit since dispersion relations are not enabled for those two CFFs.

%\section{Summary and Outlook}
{\it{Summary and Outlook.}}---
A LQCD-informed neural network extraction of six CFFs of proton DVCS is reported, allowing for a simultaneous incorporation of empirical DVCS and LQCD information.
LQCD calculations of GFFs close to pion mass is used to access the LO and NLO subtractions of dispersion relation.
%which is related to the mechanical forces acting on partons inside the nucleon \cite{Polyakov:2002yz,Polyakov:2018zvc}.
The uncertainties of CFFs ${\mathcal{H}}$ and ${\mathcal{E}}$ obtained in domains covered by available data and LQCD calculations are encouraging,
highlighting the improvements after incorporating LQCD calculations,
in particular for large $\xi$, $-t$,and $Q^2$ regime that are
difficult to access directly through DVCS experiments.
%Over a wide range of $-t$ values, the new model exhibits significantly smaller uncertainties in both the $\text{Re}\mathcal{H}$ and $\text{Re}\mathcal{E}$ parts compared to the previous model.
Our results clearly indicate that the incorporation of LQCD constraints via NLO dispersive relation plays a crucial role in enabling meaningful extractions, which were not achievable using experimental data alone. Especially the subtraction constants, due to the difficulty of directly extracting them from DVCS measurements, benefit significantly from complementary information provided by lattice QCD.

On the other hand, predicted higher-order subtractions receive no constraints from available empirical information, underscoring the crucial role of new developments in LQCD calculations of higher GPD moments \cite{Francis:2025rya} in providing further constraints for higher-order dispersion relations. Our framework opens a potential new avenue for flavor separation and gluon within NLO global analysis of DVCS \cite{Braun:2022bpn}, and and facilitates significant progress in addressing the deconvolution problem of GPDs.

%,Ji:2023xzkBraun:2020yib,Braun:2021grd,
%It is straightforward to extend the framework to NLO case \cite{Cuic:2023mki}. The Kinematic power \cite{Braun:2022qly,Braun:2025xlp} and NNLO corrections \cite{Braun:2020yib,Braun:2021grd,Braun:2022bpn,Ji:2023xzk} are interesting as well.

%the power of DR constraints is appropriate.
%dispersion-informed neural networks, lattice QCD-informed  neural networks

\bigskip

Data availability---
The code supporting the findings of this letter is openly available at
\url{https://gepard.phy.hr/}

\bigskip

\begin{acknowledgments}

We are deeply grateful to Kre\v{s}imir Kumeri\v{c}ki, whose contributions were essential to the completion of this study.
We would like to thank Dimitra A. Pefkou for sending us the correlated LQCD calculations in ref. \cite{Hackett:2023rif} and Cedric Mezrag and Yao Ji for useful discussions on dispersion relations.
This work is supported by the National Key R\&D Program of China under Grant No. 2023YFA1606703, and the National Natural Science Foundation of China  (Grants No. 12547111).
We acknowledge the support of High Performance Computing Cluster of the Southern Nuclear Science Computing Center (SNSC).

\end{acknowledgments}

\bigskip

\bibliography{refs.bib}

\bigskip

\bigskip

\beginendmatter \label{apdx:endm}

%xcff_t_xi05_Q24_KM15.pdf

\begin{figure}[H]
    \centering
    \includegraphics[width=0.98\linewidth]{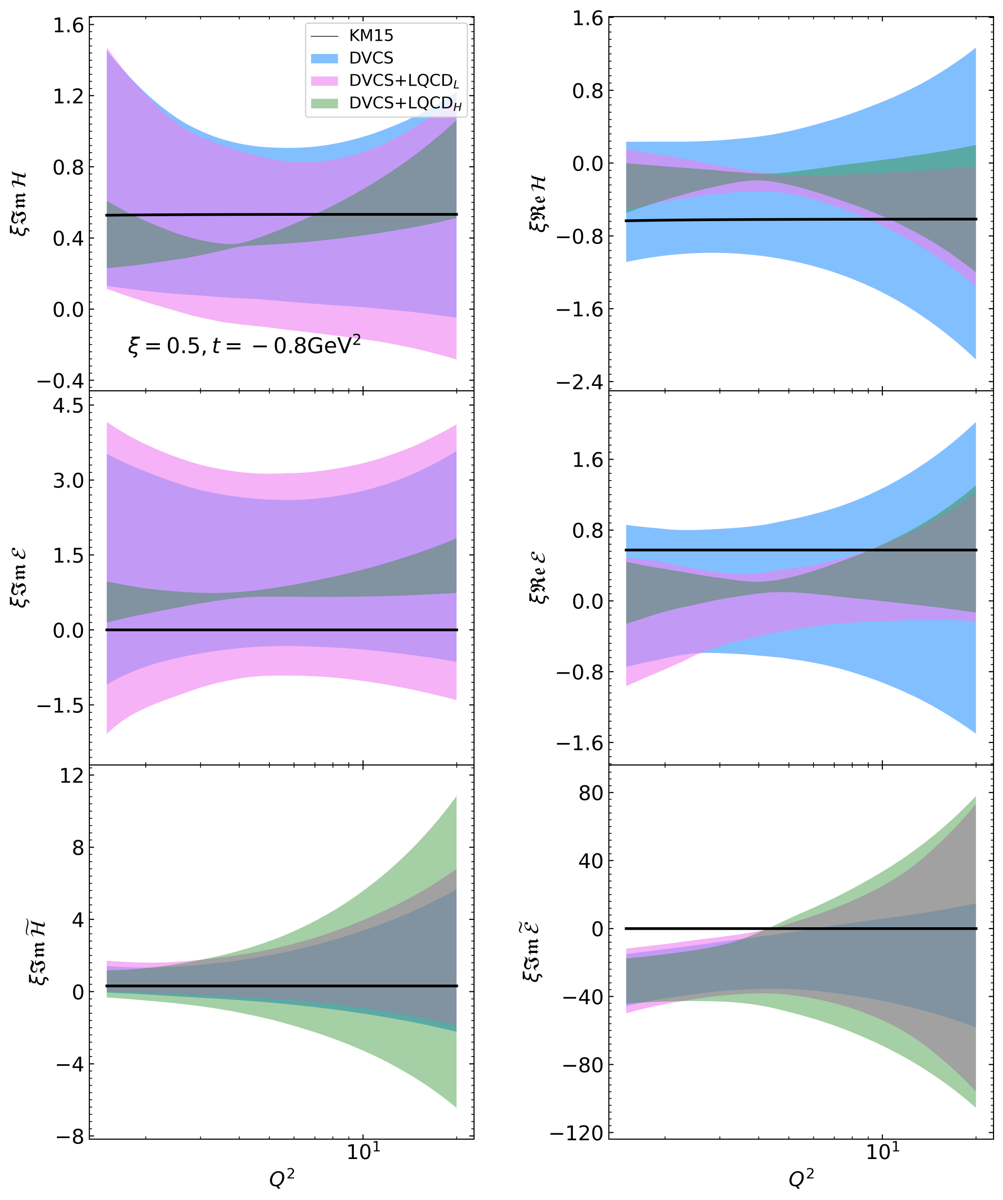}
    \caption{Imaginary and real parts of the CFFs as a function of $Q^2$ for $\xi=0.5$ and $t=-0.8$ GeV$^2$}
    \label{fig:CFFQ2}
\end{figure}
%xcff_log_Q2_xi05_t08_KM15.pdf

The $t$ and $Q^2$ dependence of CFFs are shown in Figs. \ref{fig:CFFt} and \ref{fig:CFFQ2}, respectively.

%\include{apdx}

%\newpage

\end{document}